\documentclass[a4paper,prb,twocolumn,showpacs,floatfix,footinbib]{revtex4}
\usepackage{amsmath,amssymb}
\usepackage{graphicx}

\begin{document}
\newcommand{\ds}{\displaystyle}
\newcommand{\ud}{{\mathrm{d}}}
\newcommand{\us}{{\mathrm{s}}}
\newcommand{\ubi}{{\mathrm{bi}}}
\newcommand{\LD}{L_{\mathrm{D}}}
\newcommand{\uc}{{\mathrm{c}}}
\newcommand{\ub}{{\mathrm{b}}}
\newcommand{\uB}{{\mathrm{B}}}
\newcommand{\uf}{{\mathrm{f}}}
\newcommand{\tc}{$T_c$}
\newcommand{\bd}{B33S421\,}
\newcommand{\hct}{$H_{c2}$}

\title{Resistance of superconducting nanowires connected to normal metal leads}

\author{G.R. Boogaard}
\author{A.H. Verbruggen}
\author{W. Belzig}\altaffiliation[Current
address: ]{Department of Physics and Astronomy, University of
Basel, Klingelbergstrasse 82, CH-4056 Basel, Switzerland}
\author{T.M. Klapwijk}
\affiliation{Kavli Institute of Nanoscience, Faculty of Applied
Sciences, Delft University of Technology, Lorentzweg 1, 2628 CJ
Delft, The Netherlands}

\date{\today}

\begin{abstract}
We study experimentally the low temperature resistance of
superconducting nanowires connected to normal metal reservoirs. We
find that a substantial fraction of the nanowires is resistive,
down to the lowest temperature measured, indicative of an
intrinsic boundary resistance due to the Andreev-conversion of
normal current to supercurrent. The results are successfully
analyzed in terms of the kinetic equations for diffusive
superconductors.
\end{abstract}

 \maketitle

Superconducting nanowires are resistive down to very low
temperatures due to dynamical changes of the macroscopic phase
(phase-slip). Thermally activated phase-slip (TAP) becomes more
likely for reduced cross-sectional dimensions because the
free-energy barrier scales with the area of the wire. Upon
approaching $T \rightarrow 0$, phase slip due to thermal
activation disappears and resistivity persists only by macroscopic
quantum tunneling through the free-energy barrier. These processes
have recently been studied in suspended carbon nanotubes coated
with a thin layer of a superconducting molybdenum-germanium
(MoGe-)alloy,\cite{MTNature2000,MTPRL2001} and receive increased
theoretical attention.\cite{Sachdev2004,BuchlerPRL2004} In a
separate experiment\cite{kociak01} ropes of single-walled carbon
nanotubes show signs of superconductivity, which should be
strongly influenced by phase-slips as well.

A second potential cause of low temperature resistance in
superconducting wires is the penetration of a static electric
field at normal-metal--superconductor interfaces. It reflects the
conversion of current carried by normal electrons into one carried
by Cooper-pairs via Andreev-reflection. It has been studied
extensively close to the critical temperature, where it is related
to quasiparticle charge imbalance.\cite{clarke72,hsiangclarke1980}
Although hardly experimentally studied, at very low temperatures a
similar resistive contribution is expected to be present,
reflecting a length of the order of the coherence length. Since
the coherence length is a sizable portion of the resistive length
of the nanowires it may contribute significantly to the measured
two-point resistance. Here we report experimental results on the
resistance of narrow superconducting wires connected to normal
metal leads(for short NSN). We find a strong contribution to the
resistance due to the conversion processes, which is analyzed and
understood in terms of the non-equilibrium theory for dirty
superconductors.

\begin{figure}
  \centering
  \includegraphics[width=8cm]{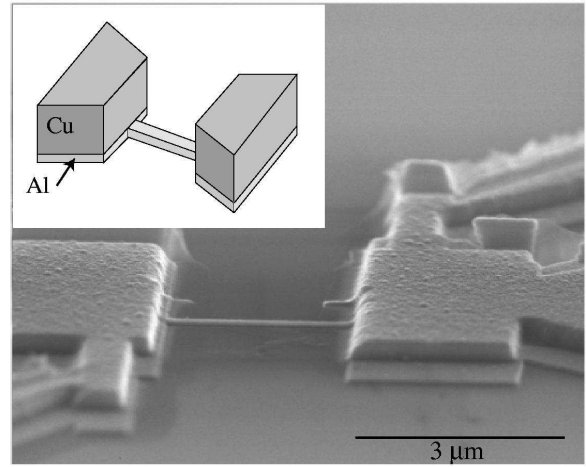}
  \caption{SEM picture of a device (slightly misaligned),
  showing the coverage of the thin aluminium film with the thick
  Cu layer (except for one the measured devices are carefully lined up).
  The inset shows a schematic picture of an ideal device.}
  \label{deduced}
\end{figure}

We chose to study samples (Fig.\ref{deduced}) made of
superconducting (S) aluminium (Al) because of its long coherence
length. Our main interest is in the two-point resistance of the
S-wire. Hence, we have chosen to work with thick and wide normal
(N) contacts with a negligible contribution to the normal state
resistance. To minimize interface resistances due to electronic
mismatch of both materials, we have chosen to work for N with
bilayers of aluminium covered with thick normal metal (Cu). In
such a geometry the superconducting aluminium wire is directly
connected to normal aluminium.

The S-wire of 100 nm thick Al is made by evaporating 99.999\%
purity Al at a rate of $\sim 1$ nm/s in a vacuum of $1\times
10^{-8}$ mbar during evaporation. Films made in the same way, have
a residual resistance ratio, $RRR = R_{300K}/R_{4.2K} $ of $7.5$
indicative of the level of impurity scattering. Taking the phonon
resistivity of $\rho_{ph}(Al) = 2.7~\mu\Omega$cm, the impurity
resistivity is $\rho_0 = 0.4~\mu\Omega$cm. Using $\sigma_0 = N(0)
e^2 D$ and renormalized free-electron parameters:  $N(0) = 2.2
\times 10^{47}$ J$^{-1}$m$^{-3}$ and $v_\mathrm{F} = 1.3\times
10^6$ m/s (Ref.~\onlinecite{ashcroftmermin}), we find the elastic
mean free path $\ell$ of 100 nm, presumably limited by the
thickness. The superconducting transition temperature of the 100
nm film is 1.26 K, the usually enhanced value for aluminium thin
films.

The sample is made in one evaporation run using shadow
evaporation. A 300 nm PMMA/ 500 nm PMMA-MAA lift-off mask defines
the width and the length of the S-wire. The width is approximately
250 nm and the length is varied from 1 $\mu$m to 4 $\mu$m. The
wire and the aluminium of the normal reservoirs is deposited in
one run, both 100 nm thick. The reservoirs are made normal by a
2$^{\textrm{nd}}$ evaporation  of 470 nm copper(Cu) on top. The Cu
is evaporated at a rate of $\sim 2.5$ nm/s at a pressure of $1
\times 10^{-7}$ mbar. The material-parameters are $\rho_0 =
0.4~\mu\Omega$cm, $\ell = 165$ nm, $D = 66\times 10^{-3}$ m$^2$/s
using $N(0) = 1.5\times 10^{47}$ J$^{-1}$m$^{-3}$, and
$v_\mathrm{F} = 1.2\times 10^6$ m/s. The non-superconducting
properties of the final reservoirs are confirmed by measuring the
resistance of a 100 nm Al/470 nm Cu bilayer down to the lowest
temperature measured: 600 mK. No sign of superconductivity is
observed. This is in agreement with the analysis of $T_c$ for a
bilayer using the Usadel equations\cite{aarts97}. Only for a
transparency of the Al/Cu interface lower than 0.1 the $T_c$ would
reach values above 600 mK. 0.1 is an unrealistically low
transparency for an in vacuum prepared Al/Cu interface.

The two-point resistance is probed with a current of 1 $\mu$A
(linear regime) in a $^3$He system down to 600 mK. The voltage
over the wire is measured with a lock-in amplifier at 133 Hz.

\begin{figure}
  \centering
  \includegraphics[width=8cm]{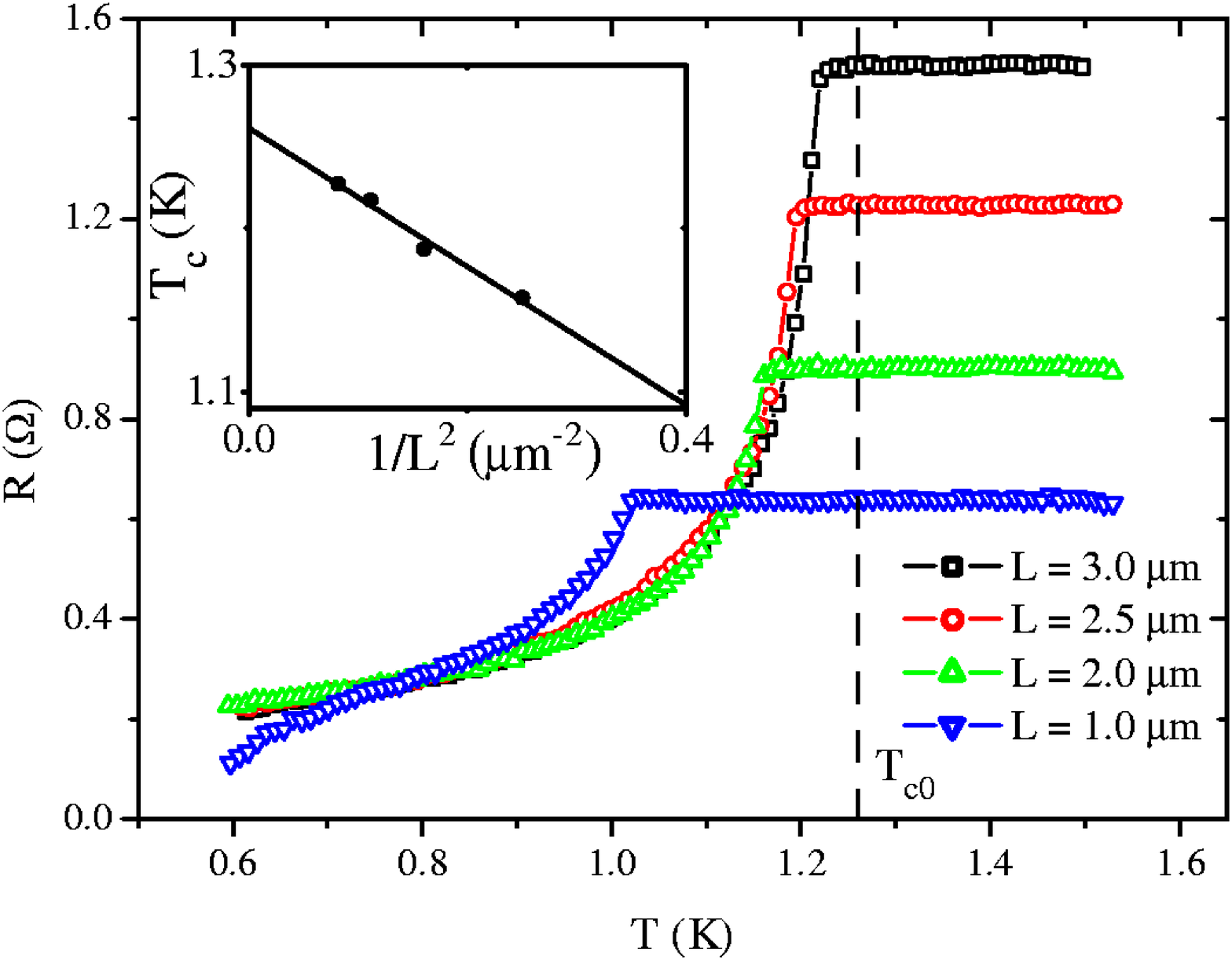}
  \caption{(color online) Measured $R-T$ curves for four different
   bridge lengths. The intrinsic $T_{c0}$ is indicated by the
   vertical dashed line. The inset shows the measured critical
   temperature of the wire \emph{vs} $1/L^2$, which is used to
   determine $T_{c0}$ by letting $L\rightarrow \infty$.
  }
  \label{realmeas}
\end{figure}

We find that the normal state resistance of nominally identical
wires scatters substantially (10\% or more), presumably due to
grain-sizes compared to wire-widths. To allow a quantitative
evaluation we have selected a set of wires with identical values
of $RRR = 3.2\pm 0.1$. Figure \ref{realmeas} shows the $R-T$
measurements for this particular set of wires. Samples with
different $RRR$ values show qualitatively identical behavior. All
wires show a finite remaining resistance as most striking
result.\cite{siddiqi00}

Obviously, despite the differences in length, the resistances at
low temperatures have identical values and follow the same trace.
It indicates that the origin of this remaining resistance is
likely due to the region in the S-wire next to the interface with
the normal reservoir. The resistance at 600 mK is equal to a
normal segment of the superconductor with a length of about 200
nm.

Figure \ref{realmeas} (Inset) also shows that the critical
temperature of the wire decreases linearly with increasing the
inverse square of the wire length.

The critical temperature of the wire should follow a
straightforward Ginzburg-Landau analysis. For
$\xi_{\mathrm{GL}}(T) \leq \pi L$ the bridge should become
superconducting. This leads to an $\sim 1/L^2$ onset-temperature
according to:
\begin{equation}
\frac{T_c}{T_{c0}} = 1 - 2.2 \pi \frac{\hbar D}{\Delta(0)}
\frac{1}{L^2}
\end{equation}
For $L \rightarrow \infty$ the normal contact can no longer
depress $T_c$ and we find the intrinsic critical temperature. For
the studied wires it is found to be $T_{c0} = 1.26$ K, identical
to the independently determined values of the 100 nm Al film.

From the normal state resistance we infer an impurity resistance
of $\rho_0 = 1.1~\mu \Omega$cm, in accordance with the $RRR =
3.2$. It is significantly higher than the $\rho_0$ of the 100 nm
Al film, leading to a lower diffusivity and elastic mean free
path: $D = 160$ cm$^2$/s, and $\ell = 37$ nm, respectively. The
resulting coherence length $\xi = \sqrt{\frac{\hbar D}{2 \pi
k_{\mathrm{B}} T_c}} = 124$ nm.

Finally, we estimate a resistance contribution of 11 m$\Omega$ due
to the spreading resistance in the normal reservoirs at low
temperatures, considerably less than the value we measure in Fig.
\ref{realmeas}.

%%%%%%%%%%%%%%%%%%%%%%THEORY DESCRIPTION%%%%%%%%%%%%%%%%%%%%%%%%%%%

Theoretically, since the studied nanowires show diffusive
transport, the Usadel equations should apply to the system.
\cite{diffusive} It is most convenient to calculate the normal
current for a given applied voltage difference, assuming linear
response. Schmid and Sch\"on\cite{schmid75} have shown that within
this limit the normal current can be described with a variation,
$\delta f(E,x)$, in the electronic distribution function $f(E,x)$:
\begin{equation}
I = \frac{A\sigma}{e} \int_{-\infty}^{+\infty}
M(E,0)\frac{\partial \delta f(E,0)}{\partial x}dE.
\end{equation}
with $x$ the coordinate along the length of the wire. $\delta
f(E,x)$ is determined from a Boltzmann-equation, which includes
the conversion of electrons into Cooper-pairs, but ignores
inelastic electron-phonon scattering (only relevant close to
$T_c$):
\begin{equation}\label{boltzmann}
\hbar D \frac{\partial}{\partial x} \left(M\frac{\partial \delta
f}{\partial x}\right) - 2\Delta N_2 \delta f = 0.
\end{equation}
The position-dependent spectral conductivity $M(E,x)$ consist of
two parts $N_1(E,x)=Re(G)$ and $N_2(E,x)=Re(F)$, with $G$ the
normal and $F$ the anomalous Greens function:
\begin{equation}
M(E,x) = \left(N_1(E,x)\right)^2 + \left(N_2(E,x)\right)^2 ,
\end{equation}
$N_1(E,x)$ is comparable to the standard BCS density of states.
The applied voltage $V$, is taken into account via the boundary
conditions for Eq. \ref{boltzmann}:
\begin{equation}
\delta f(E;0,L) = \frac{\pm eV/2}{4k_B T \cosh^2(E/2k_B T)}.
\end{equation}
Hence, the normal metal leads are taken as equilibrium reservoirs.
The strength of the pairing interaction $\Delta(x)$ in Eq.
\ref{boltzmann} is determined by solving the Usadel equation in
the Matsubara representation:
\begin{eqnarray}
\frac{1}{2} \hbar D \frac{\partial^2 \theta}{\partial x^2} &=&
-\Delta(x) \cos \theta + \omega_n \sin \theta  \nonumber \\
\omega_n &=& (2n+1) \pi k_B T,\qquad n = 0,1,\cdots
\end{eqnarray}
in conjunction with the self-consistency equation
\begin{equation}
\Delta \ln \left( \frac{T_c}{T}\right) = 2\pi k_B
T\sum_{n=0}^{\infty} \left(\frac{\Delta}{\omega_n} - \sin
(\theta)\right).
\end{equation}
The so-called proximity angle, $\theta$, parameterizes the normal
Green's function $G = \cos \theta$ and the anomalous Green's
function $F = \sin \theta$. The spectral functions are calculated
by again solving the Usadel equation but now for $\omega
\rightarrow -iE$.  The large normal metal reservoirs impose the
boundary condition $\theta(x=0,L) = 0$

Our main interest is the question how the current conversion
process contributes to the resistance. First of all, the presence
of decaying normal electron states suppresses the gap in the
density of states.

\begin{figure}
  \center
  \includegraphics[width=9cm]{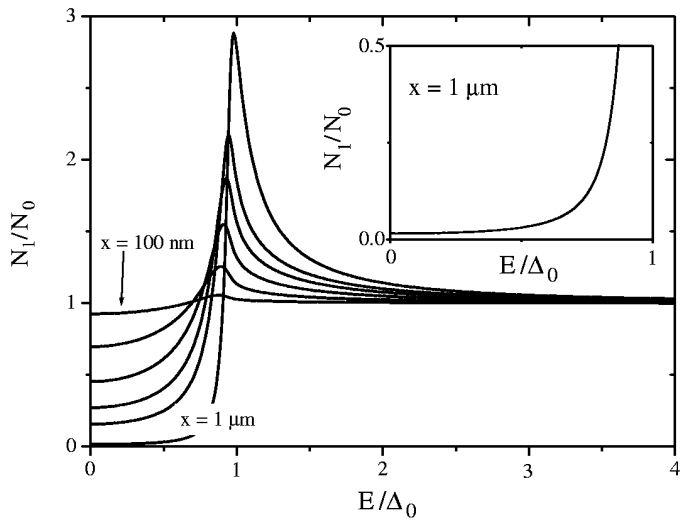}
  \caption{The calculated density-of-states $N_1$ at various distances
  from the reservoirs ($x = ~ 100, 200, 300, 400, 500, 1000~ nm$)
  for a $2~ \mu m$ long wire ($t = 0.4$,$D = 160 cm^2/s$ and
  $\Delta_0 = 192~\mu~eV$). Note the exponentially small but finite
  sub-gap density-of-states in the middle (at $x= 1 \mu m$; see inset). }
  \label{DOS}
\end{figure}

In Fig. \ref{DOS}, the density-of-states $N_1(E)$ is given for
several positions along the wire of $L = 2$ $\mu$m, $D = 160$
cm$^2$/s, $\Delta_0 = 192$ $\mu$eV, and $T/T_c = 0.4$. Clearly,
moving away form the normal leads the density of states resembles
more and more the well-known BCS density of states. Note however,
that a finite sub-gap value remains in the middle ($x= 1 \mu$m)
even for very long wires. This is an intrinsic result for any NSN
system and it is not due to current-flow, since this result is
calculated in thermal equilibrium. (The back-action of the
current-flow on the spectral properties can be neglected in the
linear response regime).

\begin{figure}
  \center
  \includegraphics[width=8.6cm]{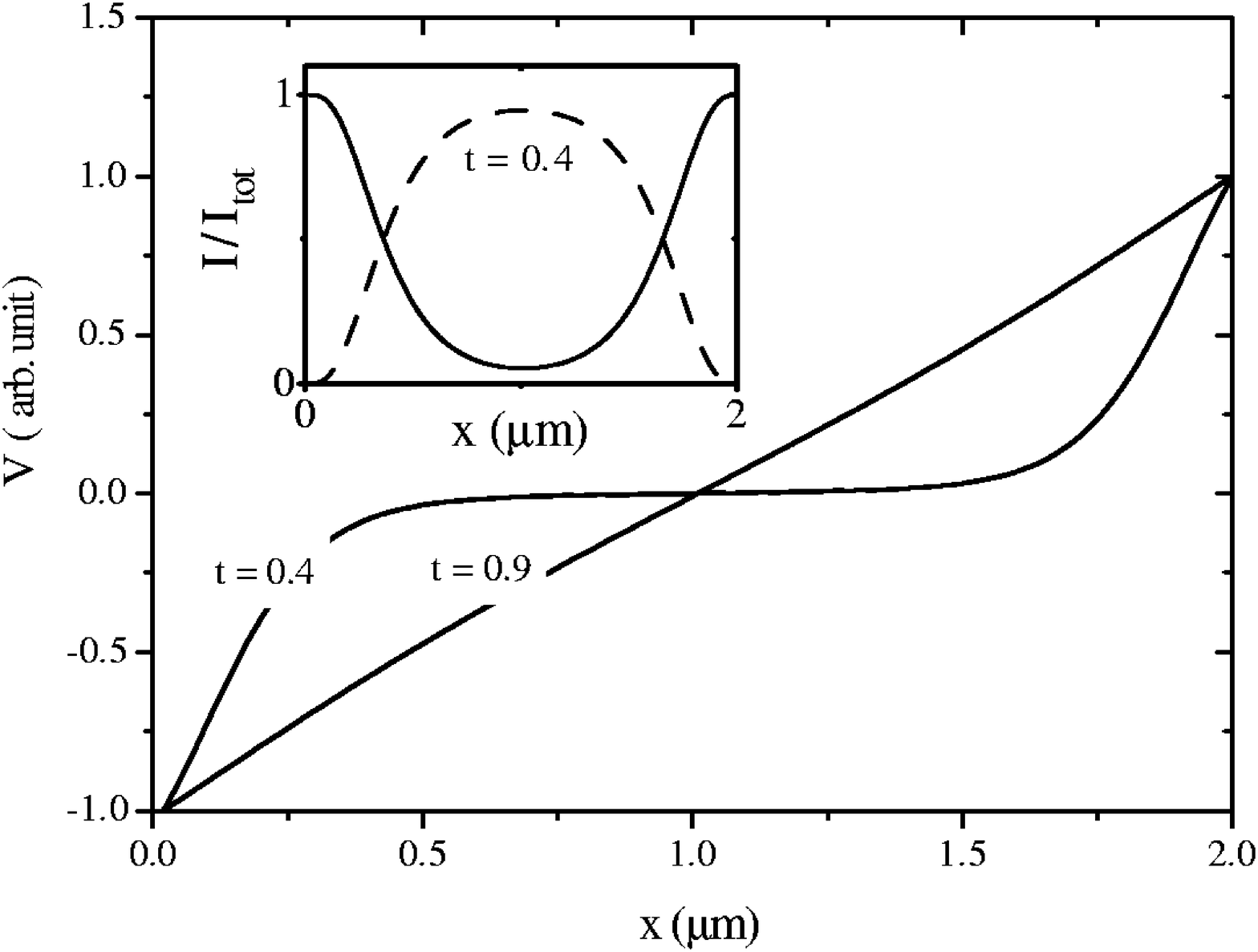}
  \caption{The voltage in the superconducting wire as a function
  of position for two different temperatures ($t=0.4$ and $0.9$).
  At $t=0.9$ the wire behaves as a normal metal and for $t=0.4$
  the voltage is clearly present to a depth $\xi$ (wire length
  2 $\mu$m with $D = 160$ cm$^2$/s and $\Delta_0 = 192$ $\mu$eV).
  The inset shows the position dependent normal currents and
  supercurrents}
  \label{electric}
\end{figure}

In Fig. \ref{electric}, we show the results of a calculation of
the voltage as a function of position along the wire for two
different temperatures: $t = 0.4$, and $t=0.9$ with $D = 160$
cm$^2$/s, and $\Delta_0 = 192$ $\mu$eV. At the temperature close
to the transition temperature, the electric field penetrates the
sample completely and the resistance is close to the normal state
value. At low temperatures, the electric field still penetrates
the superconductor over a finite length, leaving a middle piece
with hardly any voltage drop. \cite{localvoltage} The penetration
length is of the order of the coherence length. The inset shows
the position dependent normal currents and supercurrents
illustrating the current conversion processes.

\begin{figure}
  \centering
  \includegraphics[width=8cm]{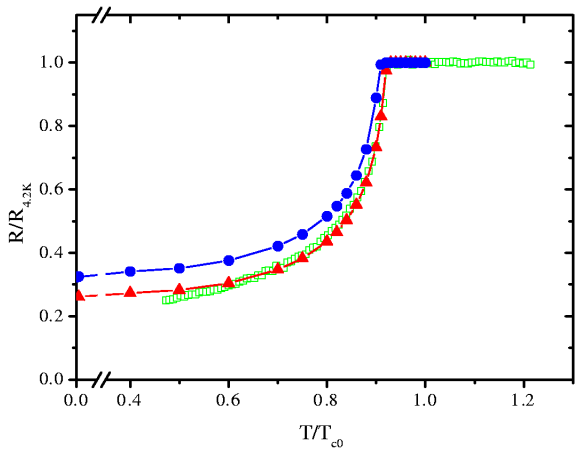}
  \caption{(color online)The measured R-T curve for the 2 $\mu$m
  bridge together with the model calculation using the boundary
  condition $\theta (x = 0,L) = 0$ (dots), and using the boundary
  condition $\partial_x \theta (x = 0,L) = \theta (x = 0,L)/a$
  (triangles) with $a = 75$ nm. The value for T=0 is 0.3255 for
  the hard boundary conditions(dots), and 0.2609 for the soft boundary
  conditions}
  \label{final}
\end{figure}

In Fig. \ref{final}, a comparison is made between the calculated
resistance as a function of temperature and the measurement for a
$L = 2$ $\mu$m wire. The calculation is done with parameters $D =
160$ cm$^2$/s, as determined from the impurity resistivity, and
$\Delta(0) = 1.764 k_{\mathrm{B}} T_c = 192$ $\mu$eV with $T_c =
1.26$ K determined from the length dependence. These parameters
have been determined independently. Without any fitting parameter,
the agreement between the model (dots) and the experiment (data
points: open symbols) is encouragingly good. Only at the lower
temperatures the observed resistance is slightly less than the
theoretically predicted values.

%%%%%%%%%%%%%%%%%%%%%%%%%%%%%%%%%%%DISCUSSION%%%%%%%%%%%%%%%%%%%%%%

Apparently, the model is overestimating the remaining resistance
below $T = 0.8$~K . Since there is little freedom left, we have
very few options to remedy this discrepancy. The most likely
option is that the rigid boundary conditions imposed at the
interfaces should be relaxed. There is a finite possibility for
superconducting correlations to extend into the normal metal
reservoirs, which would mean that the boundary condition $\theta
(x = 0,L) = 0$ is too rigid. Since the correlations will extend
into a 3-dimensional volume we assume that using the boundary
conditions $\partial_x \theta (x = 0,L) = \theta (x = 0,L)/a$
\emph{i.e.} a decay over a fixed characteristic length $a$, is a
realistic assumption. It assumes a geometric dilution of the
correlations. The result is shown in Fig. \ref{final} by filled
triangles. The best agreement between measurement and calculation
model is obtained for $a = 75$ nm. This value appears reasonable
for a decay length since it is comparable to the dimensions (100
nm by 250 nm) of the wire which emits into the reservoirs. The
shortest bridge shown in Fig. \ref{realmeas} is found to have a
further reduction in resistance, which we attribute to a small
misalignment as shown in Fig. \ref{deduced}. Note however that the
model predicts that NSN devices will continue to be resistive down
to $T = 0$ K. For the rigid boundary conditions we find
$R/R_{4.2K} = 0.3255$ and for the soft boundary conditions we find
$R/R_{4.2K} = 0.2609$

An early indication of the low temperature boundary reported here
is given by Harding \emph{et al.}\cite{harding74}. They studied
the resistance of thick sandwiches of Pb-Cu-Pb. By subtracting the
contribution to the resistance of the Cu-layers they identified a
remaining boundary resistance which depended on the mean free path
in the Pb layers. In later work by Hsiang and
Clarke\cite{hsiangclarke1980} such a  resistance appeared to be
unobservable, in contrast however to more recent work by Gu
\emph{et al.} \cite{PrattPRB2002} The geometry of our sample, with
a negligible contribution to the resistance from the reservoirs,
allows us to measure only the resistance due to the conversion
processes in the superconductor. In response to the work of
Harding \emph{et al.} Kr\"ahenb\"uhl and Watts-Tobin
\cite{krahenbuhl79} derived an analytical expression for the
effective length of the boundary resistance: $x_0 =
\sqrt{\frac{1}{6}\pi \xi_0 l}f(T)$ with $\ell$ the mean free path
for elastic scattering, $\xi_0$ the BCS coherence length and $f$ a
function of temperature. For $T=0$ K, the function $f$ is of order
1. In contrast to our model Ref.~\onlinecite{krahenbuhl79} allows
for one-dimensional diffusion in N, rather than treating N as an
equilibrium reservoir.

In conclusion, we have shown that the resistance of a
superconducting nano-wire connected to normal leads has a finite
DC resistance down to very low temperatures. The microscopic
theory describes the data very well and provide a detailed image
of the conversion from normal current into supercurrent, over a
few coherence lengths. The results emphasize the important role of
the length of the wires in relation to the nature of the
contacts.\cite{Sachdev2004} It also explains results obtained with
diffusion-cooled hot-electron bolometers in which normal leads are
used to provide rapid out-diffusion of hot electrons from a
superconducting wire\cite{siddiqi00}.

We thank the Stichting voor Fundamenteel Onderzoek der
Materie(FOM) for financial support. We thank R.S. Keizer and A.A.
Golubov for stimulating and clarifying discussions.


\begin{thebibliography}{13}

\bibitem{MTNature2000}A.~Bezryadin, C.N.~Lau, and M.~Tinkham,
Nature \textbf{404}, 971 (2000).

\bibitem{MTPRL2001}C.N.~Lau, N.~Markovic, M.~Bockrath, A.~Bezryadin,
and M.~Tinkham, Phys. Rev. Lett. \textbf{87}, 217003 (2001).

\bibitem{Sachdev2004}S.~Sachdev,P.~Warner, and M.~Troyer,
cond-mat/0402431

\bibitem{BuchlerPRL2004}H.P.~B\"uchler, V.B.~Geshkenbein, and
G.~Blatter, Phys. Rev. Lett. \textbf{92}, 067007 (2004)

\bibitem{kociak01} M.~Kociak, A.Yu.~Kasumov, S.~Gu\'eron,
B.~Reulet, I.I.~Khodos, Yu.B.~Gorbatov, V.T.~Volkov, L.~Vaccarini,
and H.~Bouchiat. Phys. Rev. Lett. \textbf{86}, 2416 (2001);
A.~Kasumov, M.~Kociak, M.~Ferrier, R.~DeBlock, S.~Gu\'eron,
B.~Reulet, I.~Khodos, O.~St\'ephan, H.~Bouchiat, Phys.Rev.
B\textbf{68}, 214521 (2003).

\bibitem{clarke72} J.~Clarke,
  Phys. Rev. Lett. \textbf{28}, 1363 (1972).

\bibitem{hsiangclarke1980}T.~Y.~Hsiang and J.~Clarke, Phys. Rev. B
\textbf{21}, 945 (1980)

\bibitem{ashcroftmermin} N.W.~Ashcroft and N.D.~Mermin, \emph{Solid
State Physics} (Saunders College, New York, 1976).

\bibitem{aarts97} J.~Aarts, J.M.E.~Geers, E.~Br\"uck,
A.A.~Golubov, and R.~Coehoorn. Phys. Rev. B \textbf{56} 2779
(1997); Appendix.

\bibitem{siddiqi00} I.~Siddiqi, A.~Verevkin, D.E.~Prober,
A.~Skalare, B.S.~Karasik, W.R.~McGrath, P.~Echternach, and
H.G.~LeDuc, IEEE Transactions on Applied Superconductivity
\textbf{11}, 958 (2001); A.H.~Verbruggen, T.M.~Klapwijk,
W.~Belzig, and J.R.~Gao, in \emph{12th International Symposium on
Space TeraHertz Technology}, edited by Imran Mehdi (San Diego,
2002), p. 42.

\bibitem{diffusive} The Usadel equations assume diffusive
scattering, whereas in our wires the variation in $RRR$ suggests
that the scattering throughout the wire might be inhomogeneous, an
effect we will further ignore.

\bibitem{schmid75} A.~Schmid and G.~Sch\"on,
  J. Low Temp. Phys. \textbf{20}, 207 (1975).


\bibitem{localvoltage} This would be the voltage measured with a normal probe,
 which exchanges quasi-particles, compared to a superconducting probe,
 which measures the chemical potential of the superconducting condensate.

\bibitem{harding74} G.R.~Harding, A.B.~Pippard, and
J.R.~Tomlinson
  Proc. Roy. Soc. Lond. \textbf{A340}, 1 (1974).


\bibitem{PrattPRB2002} J.Y.~Gu, J.A.~Caballero, R.D.~Slater,
R.~Loloee, and W.P.~Pratt, Jr. Phys. Rev. B \textbf{66}, 140507(R)
(2002)


\bibitem{krahenbuhl79} Y.~Kr\"ahenb\"uhl and R.J.~Watts-Tobin,
  J. Low Temp. Phys. \textbf{35}, 569 (1979).




\end{thebibliography}
\end{document}